\documentclass[twocolumn]{aastex7}[]

\DeclareUnicodeCharacter{02BC}{\-}
\shortauthors{Yan et al.}
\usepackage{textcomp, gensymb}
\usepackage{booktabs}
\usepackage{soul}
\usepackage{float}
\usepackage{comment}
\usepackage{array}
\usepackage{makecell}
\usepackage{amsmath}
\usepackage{multirow} 
\usepackage{subcaption}
\usepackage{graphicx}

\newcommand{\secdot}{\rlap{.}^{\rm s}}
\newcommand{\arcsecdot}{\rlap{.}^{''}}

\begin{document}
\title{Discovery of Unusual Jet Orientation Variations in the Microquasar GRS\,1915+105}

\correspondingauthor{Lang Cui}
\email{cuilang@xao.ac.cn}

\author[0009-0003-6680-1628]{Xi Yan}
\affiliation{State Key Laboratory of Radio Astronomy and Technology, Xinjiang Astronomical Observatory, CAS, 150 Science 1-Street, Urumqi 830011, China}
\email{yanxi@xao.ac.cn}

\author[0000-0003-0721-5509]{Lang Cui}
\affiliation{State Key Laboratory of Radio Astronomy and Technology, Xinjiang Astronomical Observatory, CAS, 150 Science 1-Street, Urumqi 830011, China}
\affiliation{Xinjiang Key Laboratory of Radio Astrophysics, 150 Science 1-Street, Urumqi, Xinjiang, 830011, China}
\email[]{cuilang@xao.ac.cn}

\author[0000-0001-7369-3539]{Wu Jiang}
\affiliation{Shanghai Astronomical Observatory, Chinese Academy of Sciences, 80 Nandan Road, Shanghai 200030, People's Republic of China}
\email[]{jiangwu@shao.ac.cn}

\author[0000-0002-5385-9586]{Zhen Yan}
\affiliation{Shanghai Astronomical Observatory, Chinese Academy of Sciences, 80 Nandan Road, Shanghai 200030, People's Republic of China}
\email[]{zyan@shao.ac.cn}

\author[0000-0003-3079-1889]{S\'andor Frey}
\affiliation{Konkoly Observatory, HUN-REN Research Centre for Astronomy and Earth Sciences, Konkoly Thege Mikl\'os \'ut 15-17, H-1121 Budapest, Hungary}
\affiliation{CSFK, MTA Centre of Excellence, Konkoly Thege Mikl\'os \'ut 15-17, H-1121 Budapest, Hungary}
\affiliation{Department of Astronomy, Institute of Physics and Astronomy, ELTE Eötvös Loránd University, P\'azm\'any P\'eter s\'et\'any 1/A, H-1117 Budapest, Hungary}
\email{frey.sandor@csfk.org}

\author[0000-0002-7586-5856]{Sergei Trushkin}
\affiliation{Special Astrophysical Observatory of the Russian Academy of Sciences, Nizhny Arkhyz 369167, Russia}
\email[]{sergei.trushkin@gmail.com}

\author[0000-0001-9984-127X]{Timur Mufakharov}
\affiliation{State Key Laboratory of Radio Astronomy and Technology, Xinjiang Astronomical Observatory, CAS, 150 Science 1-Street, Urumqi 830011, China}
\affiliation{Special Astrophysical Observatory of the Russian Academy of Sciences, Nizhny Arkhyz 369167, Russia}
\email[]{timur.mufakharov@gmail.com}

\author[0000-0002-1404-8924]{Ruchika Dhaka}
\affiliation{Department of Physics, IIT Kanpur, Kanpur, Uttar Pradesh 208016, India}
\affiliation{State Key Laboratory of Radio Astronomy and Technology, Xinjiang Astronomical Observatory, CAS, 150 Science 1-Street, Urumqi 830011, China}
\email[]{ruchikadhaka1997@gmail.com}

\author[0000-0003-2953-6442]{Shuangjing Xu} 
\affiliation{Korea Astronomy and Space Science Institute, 776 Daedeok-daero, Yuseong-gu, Daejeon 34055, Republic of Korea}
\affiliation{Shanghai Astronomical Observatory, Chinese Academy of Sciences, 80 Nandan Road, Shanghai 200030, People's Republic of China}
\email[]{sjxu@kasi.re.kr}

\begin{abstract}

We report large day-timescale variations in the orientation of the southeast--northwest jet in the prototype microquasar GRS\,1915+105. These results are based on three-epoch East Asia VLBI Network (EAVN) observations at 6.7\,GHz, obtained during giant radio flares in 2025 detected by the RATAN-600 monitoring program. Our observations reveal the smallest position angle (PA) of $118^\circ \pm 7^\circ$ ever measured for the jet in GRS\,1915+105, which increases to $152^\circ \pm 2^\circ$ within 37\,days. Based on the literature results, we further suggest that the jet orientation has exhibited significant variations over a PA range of $118^\circ$--$188^\circ$ since 2023. This unusual jet orientation behavior in GRS\,1915+105 during its current X-ray–obscured state may arise from a warped, precessing inner accretion disk, as implied by recent X-ray spectroscopy. Notably, one image reveals a peculiar morphology in GRS\,1915+105, which likely indicates lateral spreading of the approaching southeast jet. Future observations are essential to clarify the issues raised in this work.

\end{abstract}

\keywords{\uat{Low-mass x-ray binary stars}{939} --- \uat{Radio jets}{1347} --- \uat{Stellar mass black holes}{1611} --- \uat{Very long baseline interferometry}{1769} }

\section{Introduction} \label{sec:Introduction}
GRS\,1915+105 is a remarkable low-mass black hole X-ray binary (BHXB) in the Milky Way. Located at a distance of $9.4\pm1.0$\,kpc, it hosts a $11\pm2\,M_{\odot}$ black hole accreting from a K-type companion star \citep{Reid_2014,Reid_2023ApJ...959...85R}. This binary system is best known as the first Galactic source to display apparent superluminal motions in its radio jets \citep{Mirabel_1994Natur.371...46M,Mirabel_1999ARA&A..37..409M}. It is also the largest known X-ray binary, with a very long orbital period of $33.85\pm0.16$ days \citep{Greiner_2001Natur.414..522G,Steeghs_2013ApJ...768..185S}. Owing to these unique properties, GRS\,1915+105 has long served as a prime laboratory for studying the accretion process, relativistic jet formation, and their coupling in black hole systems \citep[e.g.,][]{Fender_2004ARA&A..42..317F}.

X-rays probe the inner regions of the accretion disk, while radio observations monitor the synchrotron emission from relativistic jets. Unlike most BHXBs, GRS\,1915+105 shows extremely complex X-ray variability including at least 12 distinct classes, which can be reduced to transitions between three basic spectral states: two soft states (A and B) and one hard state (C) \citep[e.g.,][]{Belloni_2000A&A...355..271B,Klein-Wolt_2002MNRAS.331..745K,Dhaka_2025ApJ...984..118D}. In the hard state C (or the ``plateau'' state), steady self-absorbed compact jets can be observed \citep{Dhawan_2000ApJ...543..373D,Fuchs_2003A&A...409L..35F}. During transitions from the hard to soft state, flaring optically thin ejecta are commonly captured with relativistic speeds \citep{Rodr_1999ApJ...511..398R,Fender_1999MNRAS.304..865F,Miller-Jones_2005MNRAS.363..867M,Miller-Jones_2007MNRAS.375.1087M,Rushton_2010MNRAS.401.2611R}.

Since its discovery in the 1990s, the jet of GRS\,1915+105 has generally been oriented along the southeast--northwest direction, with a mean position angle (PA) of $147^{\circ} \pm 8^{\circ}$ \citep{Rodr_2025ApJ...986..108R}. The jet viewing angle ($\theta_{\rm VA}$) has been constrained to lie between $60^{\circ}$ and $70^{\circ}$, with a weighted mean value of $64^{\circ} \pm 4^{\circ}$ \citep{Reid_2023ApJ...959...85R}. Interestingly, recent Karl G. Jansky Very Large Array (VLA) observations obtained in 2023 reveal unusual changes in both the jet PA and viewing angle relative to their historical values, by $24^\circ$ and $17^\circ$, respectively \citep{Rodr_2025ApJ...986..108R}.

Notably, after remaining active for more than 25 years, GRS\,1915+105 entered a peculiar phase in 2018, characterized by unprecedentedly low X-ray fluxes, while still exhibiting episodic radio flares and unusually bright mid-infrared emission \citep[e.g.,][]{Motta_2021MNRAS.503..152M,Gandhi_2025MNRAS.537.1385G}. This phenomenon is attributed to heavily obscuring material along the line of sight, with a highly variable column density of $N_{\rm H} \sim (10^{22}-10^{24})\,\rm cm^{-2}$ \citep[e.g.,][]{Miller_2020ApJ...904...30M,Balakrishnan_2021ApJ...909...41B,Athulya_2023MNRAS.525..489A}. Recent X-ray spectroscopic studies further suggest that this obscuration could arise from an irradiated outer accretion disk occulting the central region, possibly as a consequence of warping and precession of the inner disk \citep{Miller_2025ApJ...995L..14M}.

Despite extensive radio observations of jets in GRS\,1915+105 on (sub)arcsecond scales, very long baseline interferometry (VLBI) studies on milliarcsecond (mas) scales are still scarce \citep[e.g.,][]{Dhawan_2000ApJ...543..373D}. In this work, we present a three-epoch VLBI study of the jet in GRS\,1915+105 during its current obscured state, carried out amid multiple giant radio flares in early 2025. Section~\ref{sec:Observations} describes the observations and data reduction. The results and discussion are presented in Section~\ref{sec:Observatiosec:Results_and_Discussionns}, followed by a summary of our main findings in Section~\ref{sec:Summary}.

\begin{deluxetable*}{cccccccccc}[htbp!]
\tablecaption{Summary of EAVN Observations and Data of GRS\,1915+105 \label{table:GRS1915_observation_summary}}
\tablehead{\colhead{Epoch} & \colhead{Date} & \colhead{Stations$^a$} & \colhead{$\nu$} & \colhead{$\Delta \nu$} & \colhead{Synthesized Beam} & \colhead{$I_{\rm peak}$} & \colhead{$I_{\rm rms}$} & \colhead{$S_{\rm tot}$}\\
& & & (GHz) & (MHz) & (mas $\times$ mas, deg) & \multicolumn{2}{c}{(mJy\,beam$^{-1}$)} & (mJy) \\
(1) & (2) & (3) & (4) & (5) & (6) & (7) & (8) & (9)
}
\startdata
A & 2025/01/18 (60693.149) & EAVN ($-$VM, $-$YM) & 6.7 & 256   & $5.50\times2.60, -14.3$ & 114.6 & 0.9 & $546 \pm 109$ \\
B & 2025/01/25 (60700.170) & EAVN & 6.7 & 256  &  $5.87\times 2.87, -24.1$ & 56.2 & 0.3 & $267 \pm 53$ \\
F & 2025/02/24 (60730.049) & EAVN ($-$YM, $-$T6, $-$UR) & 6.7 & 256 & $5.56\times3.05$, $-44.0$ & 54.2 & 0.6 & $200 \pm 40$ \\
\enddata
\tablecomments{
Column~(1): observing epoch.
Column~(2): observation date and the corresponding Modified Julian Date (MJD), representing the approximate midpoint of each observation.
Column~(3): stations involved in the observations, with non-participating stations indicated by a minus sign.
At 6.7\,GHz, the EAVN consists of nine stations:  
the Ulsan (KU) station in South Korea;  
the Yamaguchi (YM)/Hitachi (HT) stations and four VERA (VLBI Exploration of Radio Astrometry) stations in Japan: Iriki (VR), Ogasawara (VO), Ishigakijima (VS), and Mizusawa (VM);
and the Shanghai Tianma (T6) and Urumqi Nanshan (UR) telescopes in China.  
Columns~(4) and (5): observing frequency and recording bandwidth. 
Column~(6): major axis position angle of the elliptical Gaussian synthesized beam, measured from north to east.
Columns~(7)--(9): peak intensity, rms noise level, and total flux density of the image, respectively (see \autoref{fig:GRS1915_light_curve_and_VLBI_maps}).
\flushleft $^a$ 
The long baselines ($\gtrsim 50\mathrm{M}\lambda$; where $\lambda$ is the observing wavelength) were strongly affected by scattering from the dense and turbulent interstellar medium, resulting in significant angular broadening. Consequently, fringes at the UR station were limited, and we excluded the UR station from the imaging process.
}
\end{deluxetable*}

\begin{deluxetable}{cccccc}[htb!]
\tablecaption{Model-Fitted Parameters for EAVN Data
\label{table:GRS1915_modelfit_full_obs_data}}
\tablehead{
\colhead{Epoch}  & \colhead{Component}  & \colhead{$r$ (mas)}  & \colhead{PA (deg)}  & \colhead{$S_{\nu}$ (mJy)} \\
& (1) & (2) & (3) & (4)
}
\startdata
   & A1 & $5.4\pm1.4$ & $151\pm15$  & $153\pm31$ \\
A  & A2 & $3.1\pm1.4$ & $-130\pm24$ & $186\pm37$  \\ 
   & A3 & $6.9\pm1.4$ & $-87\pm11$  & $207\pm41$     \\
\hline
\multirow{2}{*}{B} 
   & B1 & $7.3\pm1.5$ & $151\pm12$  & $163\pm33$   \\
   & B2 & $7.4\pm1.5$ & $-68\pm12$  & $104\pm21$  \\
\hline
\multirow{4}{*}{F} 
& F1 & $16\pm1.4$ & $143\pm5$ & $75\pm15$  \\
& F2 & $18\pm1.4$ & $159\pm4$ & $110\pm22$   \\
& F3 & $6.1\pm1.4$ & $-44\pm13$ & $4\pm1$  \\
& F4 & $16\pm1.4$ & $-36\pm5$ & $11\pm2$  \\
\hline
\enddata
\tablecomments{
Column~(1): component label (see \autoref{fig:GRS1915_light_curve_and_VLBI_maps}).
Columns~(2) and (3): distance and position angle of the component relative to the map center. We assumed that the distance uncertainty ($\sigma_r$) is one-quarter of the major-axis size of the synthesized beam and estimated the uncertainty in PA to be $\sigma_{\rm PA} = \mathrm{arctan}(\sigma_r/r)$.
Column~(4): flux density.
}
\end{deluxetable}

\section{Observations and Data Reduction} \label{sec:Observations}
In January 2025, monitoring with the RATAN-600 radio telescope (hereafter RATAN) revealed strong radio flares in GRS\,1915+105 \citep{Trushkin_2025ATel16976....1T}. Motivated by this, we promptly initiated East Asia VLBI Network (EAVN)\footnote{EAVN: \url{https://radio.kasi.re.kr/eavn/main.php}} observations at 6.7\,GHz, conducted over six epochs (A, B, C, D, E, and F; 5 hours/epoch; Project Code: A25D1) between 18 January and 24 February (\autoref{table:GRS1915_observation_summary}). 

The data were recorded in left-hand circular polarization at a rate of 1024 Mbps with 2-bit sampling, providing a total bandwidth of 256\,MHz. These data were correlated using the Daejeon correlator at the Korea--Japan Correlation Center \citep[KJCC;][]{Lee_2014AJ....147...77L}. 

Our observations involved fast switching between GRS\,1915+105 and ICRF J192540.8+122738 (J1925+1227 hereafter; which is separated by 2$\fdg$98 on the sky and almost unresolved, having a measured visibility amplitude that is almost independent of uv-distance\footnote{See \url{https://astrogeo.org/cgi-bin/imdb_get_source.csh?source_name=J1925\%2B1227}.}) with a cycle of $\sim 3$ minutes. We performed phase-referencing calibrations using the Astronomical Image Processing System \citep[{\tt AIPS};][]{Greisen2003}. A priori amplitude calibration was performed using antenna gain curves and system temperatures, with a scaling factor of 1.3 applied to account for amplitude losses in the Daejeon correlator \citep{Lee_2015JKAS...48..229L}. However, the system temperature was not measured for the YM station. Hence, we corrected its amplitude by adopting an a priori system equivalent flux density of 286\,Jy at 6.7\,GHz\footnote{See Table~16 in the EAVN Status Report for the 2026A Semester (\url{https://radio.kasi.re.kr/status_report.php?cate=EAVN}).}, using the task {\tt CLCOR}\footnote{For details, see \url{https://www.atnf.csiro.au/vlbi/dokuwiki/doku.php/lbaops/lbacalibrationnotes/pre2010?s[]=sefd}.}. Bandpass calibration was derived from scans of bright fringe-finder sources.

For phase calibration, we first read a ``recalculation table" into the data, which contains an up-to-date geodynamical model, Earth orientation parameters, station coordinates, ionospheric delay, and tropospheric delay \citep[e.g.,][]{Nagayama_2020PASJ...72...52N,Sakai_2023PASJ...75..208S}\footnote{The conventional procedure for generating the recalculation table involves KJCC providing a binary a priori delay model file, which is then used by the VERA team to recompute delays with their proprietary software. However, a recent system failure at KJCC has prevented the delivery of this file, halting the standard process. An alternative code is currently being developed by N. Sakai from National Astronomical Observatory of Japan, based on the available text-format a priori delay model files. In this work, we used recalculation tables generated by this code.}. In particular, we note that the ionospheric delay cannot be well calibrated using the recalculation tables at 6.7\,GHz-band.
We then applied corrections for the parallactic angle and instrumental delays. Following this, we carried out global fringe fitting exclusively on the calibrator J1925+1227 to derive residual delay, rate, and phase solutions. These solutions were interpolated to the scans of GRS\,1915+105. Finally, we obtained the phase-referenced images for GRS\,1915+105 in {\tt DIFMAP} \citep{Shepherd_Difmap_1997ASPC..125...77S}. We note that the data quality for epochs C, D, and E is limited, and thus, we will not present the results in this paper.

\begin{figure*}[htbp!]
\centering
\begin{minipage}[c]{0.55\textwidth}
    \centering
    \includegraphics[width=\linewidth]{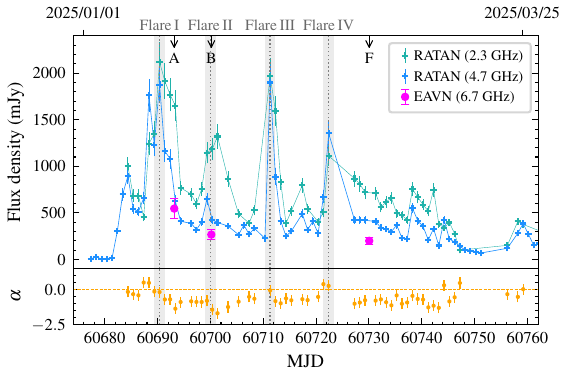}
\end{minipage} 
\begin{minipage}[c]{1\linewidth}
    \centering
    \includegraphics[width=0.37\linewidth]{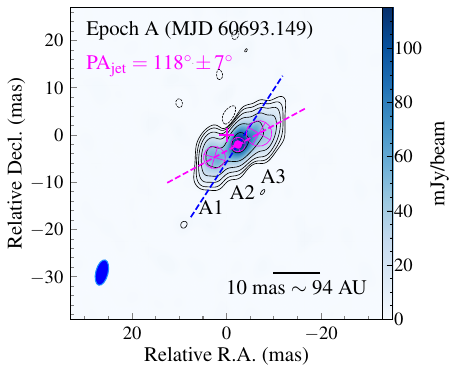}
    \includegraphics[width=0.30\linewidth]{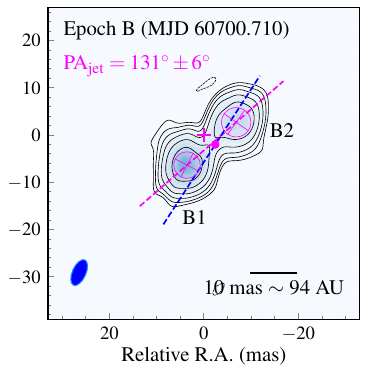}
    \includegraphics[width=0.30\linewidth]{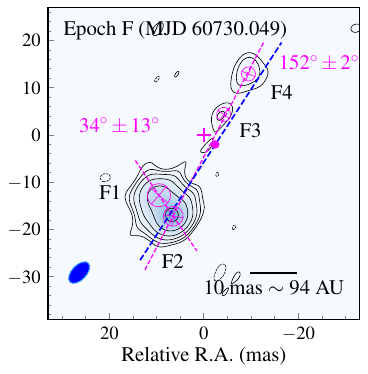}  
\end{minipage}
    \caption{
    \textit{Top:} radio light curves of GRS\,1915+105 from January to March 2025, observed with RATAN at 2.3 and 4.7\,GHz (top), together with the corresponding spectral index (bottom), defined as $\alpha = \log(S_{2}/S_{1}) / \log(\nu_{2}/\nu_{1})$. No data are available between MJD 60723 and 60726. Four radio flares are identified, peaking at MJD $60690.4 \pm 1$ (Flare~I), $60700.0\pm 1$ (Flare~II), $60711.3 \pm 1$ (Flare~III), and $60722.4 \pm 1$ (Flare~IV). Our EAVN epochs (A, B, and F) are also marked at the top of the $x$-axis, with the corresponding VLBI total flux densities overlaid. 
    \textit{Bottom:} EAVN 6.7\,GHz phase-referenced images of GRS\,1915+105 obtained in 2025. All images share the same color bar, as indicated in the epoch~A image. The FWHM of the synthesized beam is shown in the lower-left corner of each panel. Contours are drawn starting at $\pm3I_{\rm rms}$ (see \autoref{table:GRS1915_observation_summary}) and increase by successive factors of 2. Dashed contours represent negative levels. The magenta circles show the position and FWHM of the model-fitted circular Gaussian components. The magenta dashed line indicates the jet orientation (i.e., PA) derived from the component relative distances (see Section~\ref{subsec:Radio_Light_Curves}), while the blue dashed line represents the typical orientation for the southeast--northwest jet, with $\langle \mathrm{PA} \rangle = 147^\circ \pm 8^\circ$ \citep{Rodr_2025ApJ...986..108R}. The magenta cross marks the extrapolated position of GRS\,1915+105 on January 1 2025 ($19^{\rm h}15^{\rm m}11\secdot5437 \pm 1.0$\,mas, $+10^{\circ}56^{'}44\arcsecdot594 \pm 1.3$\,mas), as derived from the source coordinates on 2000 January 1 and its proper motion (see Section~\ref{subsec:Radio_Light_Curves}).
    The magenta dots correspond to the image intensity peak of epoch~A at $\sim (-2.4, -2.0)$\,mas, which is where the core is located.
    }
    \label{fig:GRS1915_light_curve_and_VLBI_maps} 
\end{figure*}

\section{Results and Discussion} \label{sec:Observatiosec:Results_and_Discussionns}

\subsection{Radio Light Curves and Jet Orientation} \label{subsec:Radio_Light_Curves}

In \autoref{fig:GRS1915_light_curve_and_VLBI_maps}, we present the radio light curves of GRS\,1915+105 observed with the RATAN \citep{Trushkin_2000A&AT...19..525T,Trushkin_2008mqw..confE..32T} at 2.3 and 4.7\,GHz from January to March 2025. Remarkably, the source exhibited multiple giant flares persisting for nearly three months, a behavior that is very exceptional. We identified four major flare events with peak flux densities occurring at MJD $60690.4 \pm 1$ (Flare~I), $60700.0 \pm 1$ (Flare~II), $60711.3 \pm 1$ (Flare~III), and $60722.4 \pm 1$ (Flare~IV). Given that the light curves were obtained from daily monitoring, we adopted a conservative uncertainty of 1 day for the flare peak times. The total flux densities measured from our EAVN observations (\autoref{table:GRS1915_observation_summary}) are also shown in \autoref{fig:GRS1915_light_curve_and_VLBI_maps}.

\autoref{fig:GRS1915_light_curve_and_VLBI_maps} also presents the EAVN phase-referenced images of GRS\,1915+105 (see \autoref{table:GRS1915_observation_summary} for the image parameters). The image center at $(0,0)$ corresponds to the extrapolated position of GRS\,1915+105 on January 1 2025 (MJD~60676), with J2000 coordinates of ($19^{\rm h}15^{\rm m}11\secdot5437 \pm 1.0$\,mas, $+10^{\circ}56^{'}44\arcsecdot594 \pm 1.3$\,mas). This position was derived from the reference position on 2000 January 1 \citep[MJD~51544; $19^{\rm h}15^{\rm m}11\secdot54902 \pm 0.7$\,mas, $+10^{\circ}56^{'}44\arcsecdot7478 \pm 0.9$\,mas;][]{Dhawan_2007ApJ...668..430D}, together with the variance-weighted average proper motion of GRS\,1915+105 reported by \citet{Reid_2023ApJ...959...85R}.

We modeled the source structure by fitting several circular Gaussian components to the visibility data using the {\tt MODELFIT} task in {\tt DIFMAP}. In this work, we are mainly interested in the distance and PA of the components relative to the map center, which allows us to derive the core position and jet orientation. In particular, we define the ``core'' to be the observed approaching jet base. Its position is frequency dependent and separated from the physical jet base (i.e., the central binary) as a result of synchrotron self-absorption. The results are summarized in \autoref{table:GRS1915_modelfit_full_obs_data} and shown in \autoref{fig:GRS1915_light_curve_and_VLBI_maps}.

As seen, in epoch~A, we detected a component (A2) near the map intensity peak at $\sim (-2.4, -2.0)$\,mas, along with two additional components on opposite sides. In epoch~B, we found that the extrapolated source position at the image center lies within the emission gap between the two bright components. These results suggest that the core is located at the map peak in epoch~A and within the emission gap in epoch~B. On the other hand, the morphology in epoch~F is dominated by a bright component in the southeast, accompanied by two components to the northwest. Notably, the bright component exhibits a northeastward extension relative to its intensity peak. We discuss this interesting morphology further in Section~\ref{subsec:epochF}.

Based on the core position and the model-fitted results in \autoref{table:GRS1915_modelfit_full_obs_data}, we derived the jet-axis PA as follows. First, we calculated the relative positions of the components in Right Ascension (R.A.) and Declination (Decl.), along with their associated errors ($\sigma_{\rm R.A.}$ and $\sigma_{\rm Decl.}$). Using the unified equations in \citet{York_2004AmJPh..72..367Y} (Eqs.~13a and 13b), we then derived the slope and intercept of the best-fit straight line, adopting an initial slope value of 1, 200 iterations, and a tolerance of $10^{-15}$. The standard errors of the slope and intercept were then estimated using their Eqs.~13c and 13d, which primarily rely on the weights for each point, defined as $w(\rm R.A.) = 1/\sigma_{\rm R.A.}^2$ and $w(\rm Decl.) = 1/\sigma_{\rm Decl.}^2$. Finally, the jet-axis PA and its uncertainty can be obtained from the slope and its standard error of the best-fit straight line, as shown in \autoref{fig:GRS1915_light_curve_and_VLBI_maps}.

In these images, we also show the typical orientation of the southeast--northwest jet, with a mean value of $\langle \mathrm{PA} \rangle = 147^{\circ} \pm 8^{\circ}$ \citep{Rodr_2025ApJ...986..108R}, assuming the core is located at $\sim (-2.4, -2.0)$\,mas. We note that epoch A reveals the smallest PA of $118^{\circ} \pm 7^{\circ}$ ever measured for the jet in GRS\,1915+105, which significantly deviates from the historical mean value by $29^{\circ} \pm 11^{\circ}$ (see also Section~\ref{subsection:Unusual_Jet_Behavior}). Moreover, epoch~B also shows a relatively small PA value of $131^{\circ} \pm 6^{\circ}$.

In epoch~F, the F2, F3, and F4 components suggest a jet-axis PA of $152^{\circ} \pm 2^{\circ}$. We note that the F1, F3, and F4 components indicate an alternate PA of $143^{\circ} \pm 2^{\circ}$ (not indicated in the image). Both values are consistent with the mean PA of the jet. In contrast, the F1 and F2 components imply that the contours of the bright component are extended along a PA of $34^{\circ} \pm 13^{\circ}$.

\subsection{The Peculiar Morphology in Epoch~F}
\label{subsec:epochF}
The morphology observed in epoch~F appears unusual compared to those in epochs~A and B, as well as to previously reported results in the literature. In this section, we discuss the epoch~F image in detail.

\subsubsection{Reliability of the F1 Component}
Is this F1 component real or an imaging artifact? To address this question, we performed a series of tests. We made images with various weighting schemes, and the F1 structure consistently appears in all cases. We also attempted iterative phase/amplitude self-calibration without including the extension in the source model. We find that the F1 component remains visible in the residual map even when the solution interval was reduced from several hundred minutes to 1 minute, yielding a reduced $\chi^{2} = 1.38$ (d.o.f = 611). By contrast, including the F1 component results in a significantly improved fit to the visibility data, with a reduced $\chi^{2} = 1.04$ (d.o.f = 589), despite using fewer degrees of freedom.

In addition, we split the calibrated data into two temporal halves and independently performed self-calibration on each subset. Both datasets revealed a consistent northeast structure to the map peak. Finally, to test whether this extension arises from any single station, we performed repeated imaging by sequentially removing individual antennas from the array. In all cases, the F1 component remains present. Based on these tests, we conclude that the F1 structure detected in epoch~F is real rather than an artifact of the imaging or calibration process.

\subsubsection{A Transversely Expanding Approaching Blob?}
We can derive the jet-to-counterjet flux density ratio ($R_{\rm S}$) using the measurements listed in \autoref{table:GRS1915_modelfit_full_obs_data}. For epoch~A, due to blending between the A3 component and the core (i.e., A2), we measured $R_{\rm S} = 0.7 \pm 0.2$, which is smaller than unity. For epoch~B, we obtained $R_{\rm S} = 1.6 \pm 0.4$. In contrast, we derived $R_{\rm S} = \mathrm{(F1 + F2)/(F3 + F4)} = 12.3 \pm 3.4$ in epoch~F, which is larger than the values measured in epochs~A and B. For comparison, previous studies have reported $R_{\rm S} < 10$ for the large-scale jet observed on scales of hundreds of milliarcseconds to arcseconds \citep[e.g.,][]{Mirabel_1994Natur.371...46M,Fender_1999MNRAS.304..865F,Miller-Jones_2005MNRAS.363..867M}. One possible interpretation is that we detected strongly asymmetric ejections in GRS\,1915+105, with the approaching blob also exhibiting lateral expansion. We note that such a peculiar jet morphology has not previously been reported in GRS\,1915+105. Future observations will be essential to clarify its origin and to investigate this phenomenon in detail.

\begin{figure*}[htbp!]
\begin{center}
    \centering
    \includegraphics[width=1\linewidth]{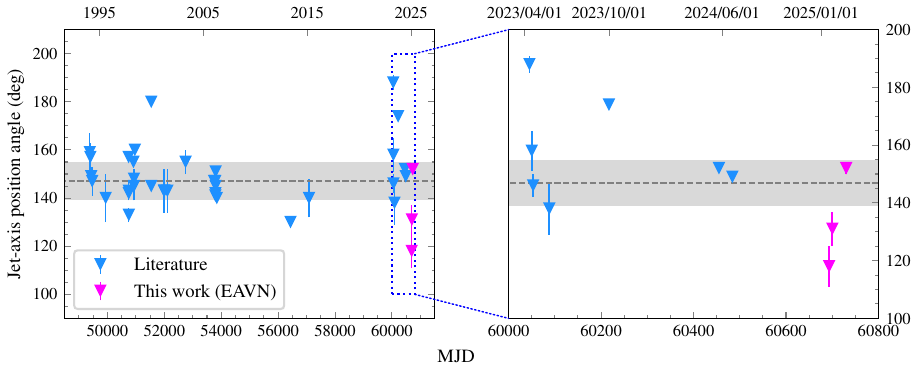}
    \caption{
    \textit{Left:} jet-axis PA of GRS\,1915+105 measured between 1994 and 2025 (see \autoref{table:GRS1915_historical_PA_Vapp_VA}).     \textit{Right:} since 2023, the orientation of the jet has exhibited significant variations over a PA range of $118^\circ$--$188^\circ$. In both panels, the horizontal dashed gray lines and shaded regions indicate the mean PA of $147^\circ \pm 8^\circ$ for the southeast--northwest jet \citep{Rodr_2025ApJ...986..108R}.
    }
    \label{fig:GRS1915_PA} 
\end{center}
\end{figure*}

\subsection{Unusual Jet Behavior of GRS\,1915+105} \label{subsection:Unusual_Jet_Behavior}
\subsubsection{Long-term Evolution of the Jet}
Our findings suggest that the jet PA varies significantly from $118^\circ \pm 7^\circ$  to $152^\circ \pm 2^\circ$ on a timescale of 37\,days.
To further study the evolution of physical properties of the jet in GRS\,1915+105 over the past three decades, we compiled key jet parameters reported in the literature, including the PA, the apparent speeds, and the jet viewing angle. As summarized in \autoref{table:GRS1915_historical_PA_Vapp_VA}, the 2023 observations suggest considerable changes in these parameters compared with their historical values: $174^\circ$--$188^\circ$ vs. $147^\circ \pm 8^\circ$ for the jet PA, $87^\circ$--$89^\circ$ vs. $64^\circ \pm 4^\circ$ for the jet viewing angle, 6.50\,mas\,d$^{-1}$/6.53\,mas\,d$^{-1}$ vs. 22.1\,mas\,d$^{-1}$/9.8\,mas\,d$^{-1}$ for the apparent speed of the approaching/receding ejecta.

In \autoref{fig:GRS1915_PA}, we show the jet-axis PA of GRS\,1915+105 measured between 1994 and 2025. The jet orientation appears to have remained relatively stable from its discovery in 1994 until 2015. Nevertheless, we note that there are a few exceptions. Early VLBI observations in 1998 by \citet{Dhawan_2000ApJ...543..373D} tentatively suggested that the innermost mas-scale jet contours were oriented at a PA of $168^{\circ} \pm 5^{\circ}$ (see their Fig.~5F). On sub-arcsecond scales, \citet{Rushton_2010MNRAS.401.2611R} found an unusual structure located to the north of the core in observations from 1999 ($\sim 180^{\circ}$; see their Fig.~3e).

In contrast, since 2023, the jet has varied significantly around $\mathrm{PA}=147^\circ$. In particular, during two different flares in 2023, VLA and VLBI observations independently detected symmetric ejecta along the south--north direction, with PAs of $174^\circ \pm 1^\circ$ and $188^\circ \pm 3^\circ$, respectively \citep{Rodr_2025ApJ...986..108R,Jiang_2026}. Notably, our image from epoch A also shows a comparable change in jet PA, but it represents the smallest PA ever observed for the GRS\,1915+105 jet ($\mathrm{PA}=118^\circ\pm7^\circ$).

\subsubsection{A Possible Warped Disk Scenario}
Large and rapid changes in jet orientation on timescales of minutes to hours have been observed in the BHXB V404~Cygni ($\Delta \mathrm{PA} \approx 36^\circ$), and are interpreted as arising from precession of the inner accretion disk \citep{Miller-Jones_2019Natur.569..374M}. Notably, these phenomena were observed during a highly obscured, super-Eddington phase of the source. A similar situation may be occurring in GRS\,1915+105, which has remained in a persistent obscured state since 2018 \citep{Motta_2021MNRAS.503..152M}. In this context, a puffed-up inner slim disk---analogous to that inferred in V404~Cygni---could form and precess as a solid body in the presence of spin--orbit misalignment, potentially giving rise to jet precession (i.e., significant changes in jet orientation).

Interestingly, recent X-ray spectroscopic observations have tentatively suggested a warped, precessing inner disk in GRS\,1915+105, which may bring the outer disk into the line of sight and naturally account for the observed obscuration and X-ray emission lines (Fig.~6 of \citealt{Miller_2025ApJ...995L..14M}). Future multi-wavelength observations will be essential to test whether this disk-jet precession scenario is indeed correct.

\begin{deluxetable*}{ccccccccccc}[htbp!]
\tabletypesize{\footnotesize}
\tablecaption{Summary of Jet Parameters in GRS\,1915+105 Measured From 1994 to 2025 \label{table:GRS1915_historical_PA_Vapp_VA}}
\tablehead{
\colhead{Date} & \colhead{Instrument} & \colhead{PA (deg)} & \colhead{$\mu_{\rm app}$} (mas\,d$^{-1}$) & \colhead{$\mu_{\rm rec}$} (mas\,d$^{-1}$)& \colhead{$\theta_{\rm VA}$ (deg)} & \colhead{Reference(s)}   \\
(1) & (2) & (3) & (4) & (5) & (6) & (7) 
}
\startdata
1994/01/29$^a$ & VLA  & $159\pm8$ & $17\pm2$ & ... & ... &  \\
1994/02/19$^a$ & VLA  & $157\pm6$ & $17.7\pm0.4$ & $7\pm2$ &  $52\pm7$ &  \citet{Mirabel_1994Natur.371...46M}\\
1994/03/19$^a$ & VLA  & $149\pm4$ & $17.5\pm0.3$ & $9.0\pm0.1$ & $64\pm2$ & \citet{Rodr_1999ApJ...511..398R}\\
1994/04/21$^a$ & VLA  & $147\pm6$ & $16.0\pm0.7$ & $8.8\pm1.0$ & $65\pm7$ & \\
1995/08/10$^a$ & VLA  & $140\pm10$ & ...  & $9\pm2$ & ... & \\ 
\hline 
1997/10/29$^a$ & MERLIN  &  $142\pm2$  & $23.6\pm0.5$ & $10.0\pm0.5$ & $62\pm2$ &  \citet{Fender_1999MNRAS.304..865F} \\
\hline 
1997/10/23$^b$ & VLBA  &  $157\pm2$ & ... & ... & ... &  \multirow{6}{*}{\citet{Dhawan_2000ApJ...543..373D}}\\
1997/10/28$^a$ & VLBA  &  $143\pm4$ & $22.1\pm1.9$$^g$ & ... & ... & \\
1997/10/31$^b$ & VLBA  &  $133\pm3$ & ... & ... & ... & \\
1998/04/11$^b$ & VLBA  & $155\pm2$ & ... & ... & ... & \\ 
1998/04/29$^a$ & VLBA  & $148\pm4$ & $22.3\pm1.7$$^g$ & ... & ... & \\
1998/05/02$^b$ & VLBA  & $145\pm6$ & ... & ... & ... & \\
\hline
1998/06/03$^c$ & MERLIN  & ... & $27\pm3$ & ... & ... & \citet{Rushton_2007MNRAS.374L..47R}\\ 
1998/05/31$^b$ & EVN  &  $\sim160$ & ... & ... & ... & \citet{Giovannini_2001ApSSS.276..111G}  \\  
1999/12/28$^b$ & MERLIN  & $\sim145$ & ... & ... & ... & \citet{Rushton_2010MNRAS.401.2611R}  \\
1999/12/28$^b$ & MERLIN  & $\sim180$ & ... & ... & ... & \citet{Rushton_2010MNRAS.401.2611R}  \\
\hline 
2001/03/21$^a$ & MERLIN  & $143\pm9$$^d$ & $20.3\pm0.7$$^h$  & $12.4\pm0.5$ & $74\pm12$ & \multirow{4}{*}{\citet{Miller-Jones_2005MNRAS.363..867M}}\\ 
2001/03/27$^a$ & MERLIN  & $143\pm9$$^d$ & $24.4\pm1.1$$^h$ & ... &... & \\ 
2001/03/31$^a$ & MERLIN  & $143\pm9$$^d$ & $25.1\pm0.7$$^h$ & ... &... & \\ 
2001/07/16$^a$ & MERLIN  & $143\pm9$$^d$ & $23.2\pm0.9$$^h$ & $12.1\pm2.0$ & $70\pm7$  \\
\hline 
2003/02/24$^c$ & MERLIN  & ... & $18\pm2$ & ... & ... &  \multirow{4}{*}{\citet{Rushton_2010MNRAS.401.2611R}} \\
2003/04/04$^c$ & MERLIN  & ... & $16.5\pm1$ & ... & ... \\
2003/05/29$^c$ & MERLIN  & ... & $17.5\pm1$ & ... & ... \\
2003/06/02$^c$ & MERLIN  & ... & $23.5\pm1$ & ... & ...  \\
\hline
2003/04/19$^b$ & VLBA  &  $155\pm5$ & ... & ... &  ...  & \citet{Ribo_2004evn..conf..111R} \\ 
\hline 
2006/01/17$^a$ & VLA   &  $147\pm3$ & $25.4\pm0.9$ & ... & ... & \multirow{7}{*}{\citet{Miller-Jones_2007MNRAS.375.1087M}} \\
2006/02/20$^a$ & VLA   &  $147\pm2$$^f$ & $17.0\pm0.2$ & ... & ... \\
2006/02/28$^b$ & VLBA &  $141\pm1$$^e$  & ... & ... & ...  \\
2006/03/04$^b$ & VLBA  &  $142\pm1$$^e$$^f$  & ... & ... & ... \\
2006/03/06$^b$ & VLBA  &  $145\pm2$$^e$  & ... & ... & ...\\
2006/03/09$^b$ & VLBA  &  $151\pm2$$^e$  & ... & ... & ...\\
\hline
2006/04/20$^b$ & EVN  & $140\pm2$ & ... & ... & ... &  \citet{Rushton_2007MNRAS.374L..47R} \\ 
2013/05/24$^b$ & VLBA  & $130\pm1$ & $23.6\pm0.5$ & ... & ... &  \citet{Reid_2014} \\ 
2015/03/05$^b$ & VLA   & $140\pm8$ & ... & ... & ... &  \citet{Rodr_2025ApJ...986..108R}  \\ 
\hline
2023/04/11$^a$  & EAVN & $188\pm3$ & $6.50\pm0.18$ & $6.53\pm0.19$ & $89\pm1$  & \multirow{3}{*}{\citet{Jiang_2026}} \\
2023/04/16$^b$ & EAVN  & $158\pm7$ & ... & ... & ... \\
2023/04/19$^b$ & EAVN  & $146\pm4$ & ... & ... & ... \\
\hline 
2023/05/23$^b$ & VLA  & $138\pm9$  & ... & ... &  ... & \multirow{4}{*}{\citet{Rodr_2025ApJ...986..108R} }   \\ 
2023/09/30$^b$ & VLA   & $174\pm1$ & ... & ... & $87\pm3$  &   \\
2024/05/25$^b$ & VLA   & $152\pm1$  & ... & ... & ... & \\
2024/06/23$^b$ & VLA   & $149\pm1$  & ... & ... &... \\
\hline
2025/01/18$^b$ & EAVN  & $118\pm7$ & ... & ... & ... & \multirow{3}{*}{This work}  \\
2025/01/25$^b$ & EAVN  & $131\pm6$ & ... & ... & ... &   \\
2025/02/24$^b$ & EAVN  & $152\pm2$ & ... & ... & ... &   \\
\enddata
\tablecomments{
Column\,(1): $^a$ ejection date; $^b$ observation date; $^c$ date of a short X-ray flare (see details in the reference).
Column\,(2): interferometer array used in the observation.
Column\,(3): position angle of the approaching ejecta or jet. 
Columns\,(4)--(5): apparent speeds of the approaching and receding ejecta, respectively.
Column\,(6): jet viewing angle, derived from columns (4) and (5) using Eq.4 in \citet{Fender_1999MNRAS.304..865F}, assuming a source distance of 9.4\,kpc.
}
$^d$ {The images shown in \citet{Miller-Jones_2005MNRAS.363..867M} have been rotated clockwise by $53^\circ$, implying a PA of $\sim143^\circ$.}
$^e$ The results are from 8.4\,GHz VLBA observations in \citet{Miller-Jones_2007MNRAS.375.1087M}.
$^f${Mean PA derived from multi-epoch measurements. }
$^g${Converted from the proper motions per hour reported in \citet{Dhawan_2000ApJ...543..373D}. }
$^h${Adopted from Table~1 in \citet{Miller-Jones_2007MNRAS.375.1087M}. }
\end{deluxetable*}

\section{Summary} \label{sec:Summary}
In 2025, our RATAN monitoring of GRS\,1915+105 detected multiple giant radio flares between January and March. Motivated by this, we conducted multi-epoch EAVN observations at 6.7\,GHz to probe the jet properties of this microquasar during its now-persistent X-ray-obscured state. Below, we summarize the main findings.

Our observations reveal the smallest PA of $118^\circ \pm 7^\circ$ ever measured for the GRS\,1915+105 jet, which increases to $152^\circ \pm 2^\circ$ within 37\,days. By combining our results with those from the literature, we suggest that the orientation of the jet remained relatively stable from its discovery in 1994 until 2015. However, since 2023, its orientation has exhibited significant variations over a PA range of $118^\circ$--$188^\circ$.

In view of discoveries in V404~Cygni \citep{Miller-Jones_2019Natur.569..374M}, we speculate that such jet PA variation may arise from the precession of the inner accretion disk in the presence of spin--orbit misalignment, which in turn could drive jet precession. Notably, a warped, precessing inner disk has been proposed recently for GRS\,1915+105 based on X-ray spectroscopic observations \citep{Miller_2025ApJ...995L..14M}. 

On the other hand, one of our images reveals a peculiar morphology in GRS\,1915+105, which may reflect highly asymmetric ejections, with the approaching component also exhibiting significant transverse expansion. Future observations are essential to investigate this phenomenon in detail.

\begin{acknowledgments}

We sincerely thank Prof.~James Miller-Jones and the referee for carefully reading the manuscript and for their constructive and insightful comments, which greatly improved the quality and clarity of the manuscript. We sincerely thank Dr.~Nobuyuki Sakai for providing the recalculation tables, which significantly facilitated this work. We thank Dr.~Sara E. Motta and Dr.~Pikky Atri for useful discussion. This work was supported by the National Key R\&D Program of China (grant Nos. 2024YFA1611500 and 2022SKA0120102). X.Y. is supported by the China Postdoctoral Science Foundation under Grant Number 2025M773200. X.Y. also acknowledges supports from the Xinjiang Tianchi Talent Program and the 2025 Outstanding Postdoctoral Grant of the Xinjiang Uygur Autonomous Region. L.C. acknowledges support from the Tianshan Talent Training Program (grant No. 2023TSYCCX0099). 
W.J. is supported by the National Natural Science Foundation of China (grant Nos. 12173074, 12573100). This work was partly supported by the Urumqi Nanshan Astronomy and Deep Space Exploration Observation and Research Station of Xinjiang (XJYWZ2303) and the Central Guidance for Local Science and Technology Development Fund (grant No. ZYYD2026JD01). 

Observations with the RATAN-600 telescope are supported by the Ministry of Science and Higher Education of the Russian Federation. The renovation of telescope equipment is currently provided within the national project ``Science and universities".

The authors sincerely thank the EAVN coordinator, Dr.~Kiyoaki Wajima, for the prompt handling of our Target-of-Opportunity requests, as well as all EAVN staff members who rapidly initiated the follow-up observations and correlated the data after the 2025 radio flares were reported. 

This work has made use of the East Asian VLBI Network (EAVN), which is operated under cooperative agreement by National Astronomical Observatory of Japan (NAOJ), Korea Astronomy and Space Science Institute (KASI), Shanghai Astronomical Observatory (SHAO), Xinjiang Astronomical Observatory (XAO), Yunnan Astronomical Observatory (YNAO), National Astronomical Research Institute of Thailand (Public Organization) (NARIT), and National Geographic Information Institute (NGII), with the operational support by Ibaraki University (for the operation of Hitachi 32-m and Takahagi 32-m telescopes), Yamaguchi University (for the operation of Yamaguchi 32-m telescope), and Kagoshima University (for the operation of VERA Iriki antenna). 
\end{acknowledgments}

\bibliographystyle{aasjournal}
\bibliography{references}

\end{document}